# High-dose high-temperature emission of LiF:Mg,Cu,P: thermally and radiation induced loss & recovery of its sensitivity

Barbara Obryk[a*], Katarzyna Skowrońska[b], Anna Sas-Bieniarz[a], Liliana Stolarczyk[a], Paweł Bilski[a]

[a]Institute of Nuclear Physics (IFJ), ul. Radzikowskiego 152, 31-342 Kraków, Poland

[b]AGH University of Science and Technology (AGH), Al. Mickiewicza 30, 30-059 Kraków, Poland

**HIGHLIGHTS:**

• High-dose high-temperature TL emission is present in LiF:Mg,Cu,P' glowcurves.

• LiF:Mg,Cu,P can measure doses ranging from below 1 µGy to about 1 MGy.

• Thermally-induced sensitivity loss of the samples can be recovered.

• Sensitivity damage of the samples after high-dose measurements is Fuldy reversible.

• High-dose measurements changes to the structure of the material are reversible.

**Abstract**

Highly sensitive LiF:Mg,Cu,P (MCP) detectors enable measurements of radiation doses from tens of nanograys up to a few kilograys, where the saturation of the signal of the main dosimetric peak occurs. Thanks to the recently observed high-dose high-temperature emission of MCP detectors heated to temperatures up to 600°C after exposures to radiation doses ranging from 1 kGy to 1 MGy, a new method of TL measurement of radiation doses has been recently developed at the Institute of Nuclear Physics (IFJ). This method can measure doses ranging from micrograys up to a megagray. So far, high dose measurements were performed on fresh MCP samples and each detector was used only once, because as a result of these measurements, the detectors lose their sensitivity to a large extent. In this study, a specific thermal treatment intended to fully restore the loss of MCPs TL sensitivity was sought. We have investigated several annealing procedures, applying different temperatures (from 400°C up to 700°C) for different periods of time (10-30 minutes) in argon atmosphere. In this way we were able to recover MCP sensitivity fully, allowing for reuse of the samples after high-dose irradiation and high-temperature measurement.

## 1. Introduction

One of the well known advantages of thermoluminescent (TL) detectors made of lithium fluoride doped with magnesium, copper and phosphorus is their very high sensitivity to ionizing radiation (Horowitz, 1993; Bos, 2001; Bilski, 2002). LiF:Mg,Cu,P (MCP) detectors enable measurements of radiation doses from tens of nanograys up to a few kilograys, when the total saturation of the signal of the so-called main dosimetric peak (at about 220°C) occurs (McKeever et al., 1995).

Only recently, at the Institute of Nuclear Physics (IFJ), the quite unexpected properties of MCP detectors at high (Bilski et al, 2007) and ultra-high doses (Bilski et al., 2008b; Obryk et al., 2009) have been observed. Significant changes of the glow-curve shape occur for doses higher than a few kilograys and the most important finding is that a new, intense peak appears for doses above 30 kGy (see Fig. 1). This peak, denoted as peak 'B', is well separated from the rest of the glow-curve and located at temperatures exceeding 400°C. After this discovery, comprehensive measurements of MCP detectors response to high doses of various radiation types: e.g. photon (Obryk et al., 2009), electron (Bilski et al., 2010), proton (Obryk et al., 2010a) and thermal and epithermal neutron (Obryk et al, 2011a), were completed (see Table 1). All the results showed the presence of the peak 'B' in the glow-curves of detectors exposed to radiation doses higher than 30 kGy.

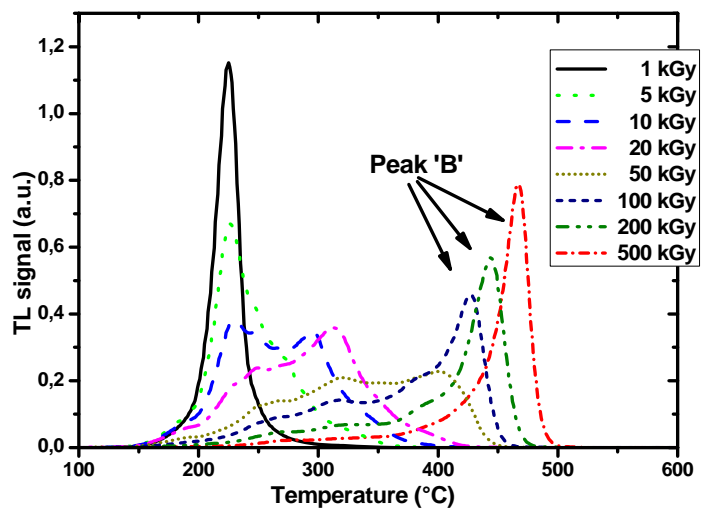

Fig. 1. MCP glow-curves resulting from gamma irradiation ($^{60}$Co source at KAERI) for the dose range 1 kGy-500 kGy.

Table 1. Qualities, energies and dose ranges of
radiation used for tests of high-dose high-temperature
emission of LiF:Mg,Cu,P detectors

| Radiation type | Radiation energy | Dose/Fluence range | Reference |
|---|---|---|---|
| Gamma | 1.25 MeV | 1 Gy - 1 MGy | Bilski et al. 2007, 2008b; Obryk et al,. 2009 |
| Electron | 6 MeV, 10 MeV | 5 kGy - 1 MGy | Bilski et al. 2010; unpublished results |
| Proton | 25 MeV, 24 GeV/c | 1 Gy - 1 MGy | Obryk et al. 2009; Obryk et al. 2010a |
| Neutron | Thermal & epithermal | $3 \times 10^{11} - 3 \times 10^{15}$ n/cm$^2$ | Obryk et al. 2011a |

Although high-dose high-temperature emission was found to be present in MCP' glow-curves after irradiation with all radiation qualities, also following irradiation in mixed fields, the mechanism leading to the peak 'B' occurrence is still under investigation. A recent study of peak 'B' spectra is analysed by Gieszczyk et al. in this issue.

On the basis of this newly discovered behaviour of MCP at high and ultra-high doses, a method of TL measurement of radiation doses ranging from micrograys up to a megagray has been recently developed at the IFJ (Obryk et al., 2010b, 2011b; Obryk, 2012). For dose measurement up to 1 kGy we exploit routine TL dosimetry which uses main dosimetric peak (peak 4) intensity, applying experimentally determined correction functions in the region of its sublinearity (Obryk et al., 2008). For doses higher than 1 kGy where significant changes of the glow-curve shape were observed the method is based on the percentage of TL signal in different temperature ranges. A parameter called the Ultra-High Temperature Ratio (UHTR) was defined in order to quantify the observed changes of the MCP glow-curve shape at very high doses and very high temperatures, which allows measuring the absorbed dose in the range from 1 kGy to 1 MGy (Bilski et al., 2010; Obryk et al., 2010a). This dosimetric method was tested in a range of radiation qualities, such as gamma radiation, electron and proton beams, thermal neutron fields and high-energy mixed fields around the Super Proton Synchrotron (SPS) and the Proton Synchrotron (PS) accelerators at CERN (the European Organization for Nuclear Research) (Obryk et al., 2008, 2011b). This method allows for ultra-high dose range (at least twelve orders of magnitude) measurement with a single MCP detector. It can be used for dosimetry at high energy accelerators, thermonuclear fusion technology facilities and has great potential for accident dosimetry in particular. A number of dosimetric sets with TL detectors are currently used around the Large Hadron Collider (LHC) at CERN.

So far, high dose measurements were performed on fresh samples and each detector was used only once, because the detectors lose their sensitivity to a large extent as a result of these measurements. The main reason is the well-known feature of LiF:Mg,Cu,P detectors: sensitivity loss when heated beyond about 270°C (Oster et al. 1993, 1996, Bilski et al.,1997, Ben-Amar et al., 1999). It was also reported by Meijvogel and Bos (1995) that high temperature readout is causing both reversible and irreversible changes of the sensitivity of the material. Also, high doses of radiation usually have some influence on various properties of the samples, among them their sensitivity (Cai et al., 1994, Bilski et al., 2008a). Following high-dose high-temperature measurements these two effects are entangled.

In this study, a specific thermal treatment intended to fully restore the loss of MCPs TL sensitivity was sought. We were encouraged by our preliminary observation that readout up to 600°C caused smaller loss of samples' sensitivity than readout up to 400°C, which is generally in accordance with the results of Tang et al. (2000) who showed that the change in glow-curve structure and the loss of TL sensitivity, as a consequence of annealing between 260°C and 400°C, can be recovered fully by annealing at 720°C for 30 min in a nitrogen atmosphere followed by the standard anneal of 10 min at 240°C.

## 2. Materials and methods

Over one thousand of highly sensitive MCP detectors, made using the sintering technique were prepared for the tests. All detectors used were of typical size: 4.5 mm diameter and 0.9 mm thickness. Detectors were produced at the Department of Radiation Physics and Dosimetry of the IFJ and belonged to one production batch with TL sensitivity homogeneity below 5%.

At each stage of the experiment the standard annealing cycle was applied: for freshly produced detectors two-phase heat treatment - 260°C for 10 minutes followed by 240°C for 10 minutes, with fast cooling using a thick metal block after each phase, while for detectors used previously one-phase heat treatment only - 240°C for 10 minutes followed by fast cooling. The individual response factor (IRF) has been also determined for each freshly produced detector using its total TL response. Irradiations have been carried out at the Secondary Standards Dosimetry Laboratory (SSDL) of the IFJ with Cs-137 source.

All readouts were performed using the MICROLAB Manual Reader-Analyzer RA'94, with a bialkali photomultiplier tube and violet filter (BG-12). Detectors were readout in argon atmosphere with a linear heating rate of 2 K s$^{-1}$. All high-temperature annealing were also carried out in argon atmosphere to avoid oxidizing of detectors' surface.

To quantify the observed changes of the glow-curve two methods of analysis have been applied. The first one consists of analysis of the main peak height, the main peak signal integral in the range of 200-230°C (integral 1), integral of the 'tail' of the glow-curve calculated in the range of 230-275°C (integral 2), the total TL signal (for the range 150-275°C), and also the ratio of integral 1 to integral 2 which we have called 'shape' parameter. The second method applied was deconvolution of all resulted glow-curves into five peaks by the GlowFit (Puchalska & Bilski, 2006) and analysis of behaviour of individual peaks. All parameters used for sensitivity analysis are presented relative to the parameters for the reference group treated routinely, i.e. as ratio of their mean values calculated for each group of detectors treated in different way during experiments to that of reference group.

## 3. Experimental

### 3.1. Thermally-induced sensitivity loss and recovery experiments

In the first stage of the experiments (1) a set of 195 detectors was prepared in order to study thermally-induced sensitivity loss of MCP detectors. All detectors were annealed in the standard way, irradiated with 5 mGy, divided into nine groups (8x20 pieces and reference group of 35 pieces) and each group was readout up to a different maximum temperature: 275 (reference group), 300, 325, 350, 400, 450, 500, 550, 600°C. Then all detectors were annealed in the standard way, irradiated (5 mGy) and readout up to 275°C.

The next stages of experimental procedure to study the recovery of thermally-induced sensitivity loss of samples were as follows (general scheme is presented in Fig. 2):

(2) 195 detectors from the previous experiment were annealed at 240°C for 30 minutes, irradiated and readout up to 275°C.

(3) All groups of detectors except the reference group were divided into four subgroups and subjected to high-temperature in gas (argon) atmosphere (HTGA) annealing: 400, 500, 600, 700°C, for 30 minutes, followed by 240°C/10 min., irradiated (5 mGy) and readout up to 275°C.

(4) The next set of detectors (195 pieces) was prepared, annealed in the standard way, irradiated with 5 mGy, divided into five groups (4x45 pieces and reference with 15 pieces) and each group was readout up to a different maximum temperature: 275 (ref. group), 300, 400, 500, 550°C.

(5) All groups of detectors except the reference group were divided into subgroups and subjected to HTGA annealing: 580, 600, 620, 640, 660°C, for 10 or 20 or 30 minutes, followed by 240°C/10 min., irradiated (5 mGy) and readout up to 275°C.

(6) The next set of detectors (195 pieces) was prepared, annealed in the standard way, irradiated with 5 mGy, divided into six groups (5x36 pieces and reference with 15 pieces) and each group was readout up to a different maximum temperature: 275 (ref. group), 300, 400, 500, 550, 600°C.

(7) All groups of detectors except the reference group were divided into nine subgroups and subjected to HTGA annealing: 600, 620, 640°C, for 10 or 20 or 30 minutes, followed by 270°C/10 min and 240°C/10 min., irradiated (5 mGy) and readout up to 275°C.

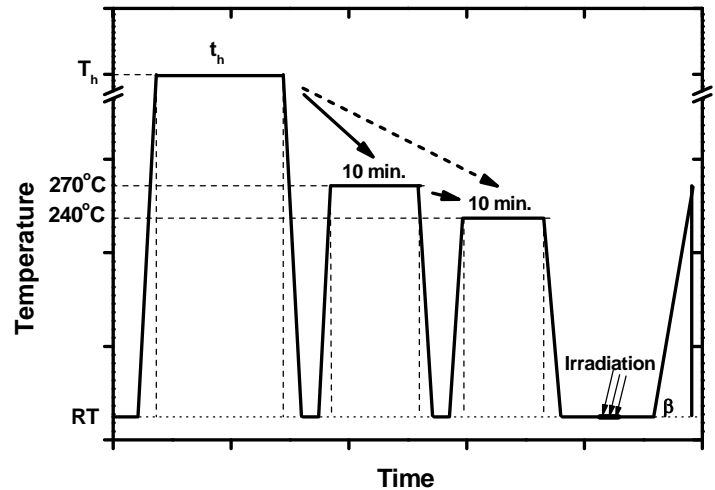

Fig. 2. Schematic representation of the investigated thermal treatment of MCP samples. The effect of different high temperatures $T_h$ of annealing in argon atmosphere followed by 270°C/10 min. and/or 240°C/10 min. on the TL response has been studied. The time $t_h$ was chosen at 10, 20 or 30 min. Cooling was immediate between metal pieces cooled earlier to the temperature -17°C in a freezer. All samples were readout at a constant heating rate $\beta=2$°C s$^{-1}$.

After high-temperature annealing detectors were cooled immediately by putting them between metal pieces which had been cooled earlier to the temperature -17°C in a freezer.

### 3.2. Radiation-induced sensitivity loss and recovery experiments

To study the recovery of radiation-induced sensitivity loss of samples two groups of detectors from our previous high-dose experiments were used:

I. detectors (about 100 pieces) previously irradiated with nine steps of doses in the range 0.5 – 500 kGy of $^{60}$Co gammas at the Korea Atomic Energy Research Institute (KAERI) and readout up to 600°C (Bilski et al. 2008b);

II. detectors (15 pieces) previously irradiated with three steps of doses: 5, 20 and 100 kGy of 10 MeV electrons at the Institute of Nuclear Chemistry and Technology in Warsaw, Poland and readout up to 550°C (Bilski et al. 2010) and five not used detectors from the same production batch.

The experimental procedure consists of standard annealing, 50 mGy irradiation and readout of all samples up to 275°C, then the developed HTGA annealing procedure was used with different temperature intervals. This procedure consists of HTGA annealing at 620°C, followed by 270°C/10 min and 240°C/10 min, 50 mGy irradiation and readout up to 275°C. For group I of detectors three different intervals of 20, 30 and 60 minutes of the 620°C anneal were applied consecutively, while for group II of detectors only 60 minutes of the 620°C anneal was applied.

## 4. Results and discussion

### 4.1. Thermally-induced sensitivity loss

The typical glow-curves resulting from the thermally-induced sensitivity loss experiment (stage (1)) are presented in Fig. 3. The shape of the resulting glow-curves significantly depends on the maximum temperature of the previous readout. The maximum of the main dosimetric peak (peak 4) decreases for increasing previous readout's temperature up to 500°C, then increases. The 'tail' of the glow-curve grows for increasing previous readout's temperature up to 550°C, then shrinks back.

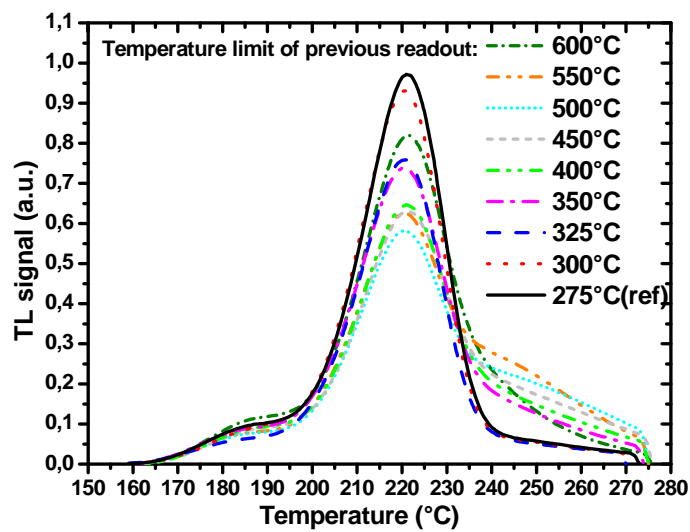

Fig. 3. The typical glow-curves of detectors previously readout up to different maximum temperatures in the range 275°C to 600°C.

Fig. 4 presents quantitatively the sensitivity loss of samples as a result of readouts up to different temperature limits higher than recommended for this material. It turned out that the total TL signal parameter is not useful for our purposes (to recover not only TL signal of the main dosimetric peak but also to regain the original shape of the glow-curve) as the decrease of the main peak height was compensated for by the growing 'tail' of the glow-curve for growing temperatures of annealing. Other parameters proved to be more suitable for our purposes and we use them in further analysis.

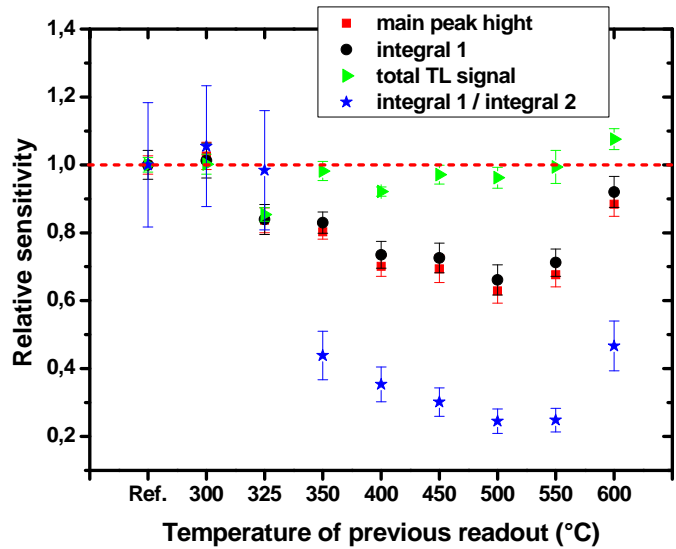

Fig. 4. Results of the multi-parameter analysis of relative sensitivity of detectors readout previously up to different temperature limits.

**4.2. Effect of high-temperature annealing on thermally-induced sensitivity recovery**

A certain sensitivity recovery of MCP detectors was achieved by prolongation of annealing at routine temperature of 240°C to 30 minutes (stage (2)) but results were not satisfactory. Resulting qualitative changes of the glow-curve after applying 30 minutes of high-temperature annealing: 400, 500, 600, 700°C in argon atmosphere followed by 240°C for 10 min. (stage (3)) are presented in Fig. 5. It turned out that the annealing temperatures 400°C and 500°C definitely change the structure of the glow-curve, thus significantly worsening the sensitivity of detectors. The intensity of the high-temperature 'tail' of the main peak becomes comparable to intensity of the peak itself, and for 500°C is even higher; also the number of counts is relatively small. However, glow-curves for 600°C and 700°C are similar to that of a typical LiF:Mg,Cu,P glow-curve.

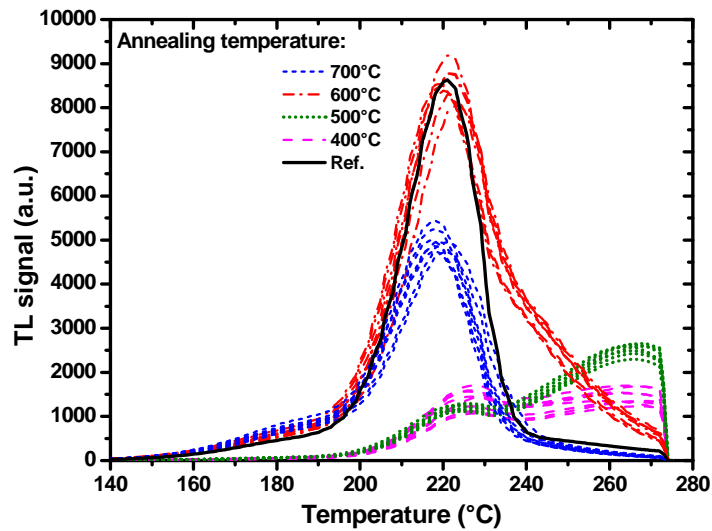

Fig. 5. The sets of glow-curves after HTGA annealing at temperatures: 400, 500, 600, 700°C in comparison to reference glow-curve (each set consists of eight typical glow-curves of detectors' groups readout previously up to different maximum temperatures in the range 275°C to 600°C).

Fig. 6 compares mean values of two parameters of quantitative analysis of these results for detectors annealed at different temperatures. Annealing at 600°C appears more suitable for the recovery of the main peak height (and also area under the peak) while taking into account the complete recovery of the original glow-curve shape far better temperature is 700°C.

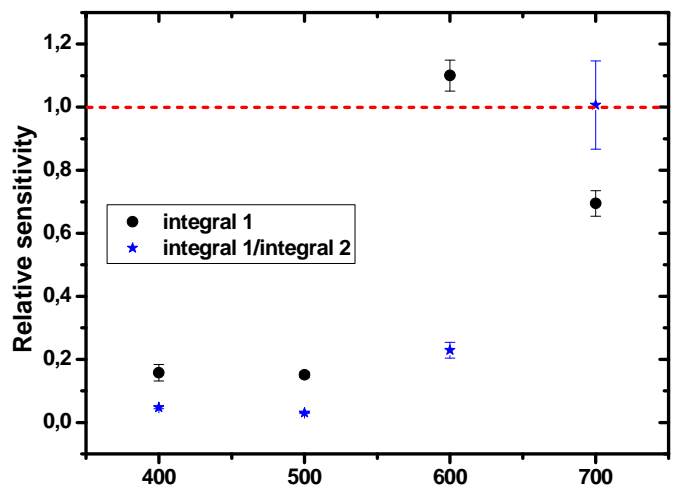

Fig. 6. Relative sensitivity of detectors presented as the mean value for all groups of detectors annealed for 30 minutes at the same temperature (in the range 400-700°C).

It was also found that the use of IRFs of detectors increases scatter of above presented results. It turned out that dispersion of results shown in Fig. 6 calculated with IRFs compared to these calculated without them increases with increasing temperature of HTGA annealing as presented in Table 2. Thus we concluded that IRFs evaluated before high-temperature treatment are rather useless after high-dose and high-temperature TL measurement. Due to this all presented results are those calculated without using IRFs.

Table 2. Ratio of standard deviation (SD) of results obtained with and without IRF after HTGA annealing

| HTGA annealing temperature [°C] | $SD_{IRF}/SD_{noIRF}$ |
|---|---|
| 400 | 0.98±0.05 |
| 500 | 1.01±0.08 |
| 600 | 1.23±0.11 |
| 700 | 1.28±0.09 |

Fig. 7 shows the typical glow-curves obtained for each annealing temperature from the range 580°C-660°C (stage (5)) for all annealing times applied and for all detectors groups readout previously to different maximum temperatures (in view of their small differences compared to variation). The glow-curves obtained are much more similar to the glow-curves of the reference group than those obtained in the previous experiment (stage (3)). It can be seen that in terms of the main dosimetric peak (peak 4) intensity, the closest to the reference group were sets of detectors annealed at 600°C and 620°C, and the higher the annealing temperature the better the reduction of the high temperature 'tail' of the main peak (peak 5). It was also noted that for most annealing temperatures the area under peak 3 was increased and a reduction of the area under peak 2 was also observed. Peak 1 was not observed even for the reference group.

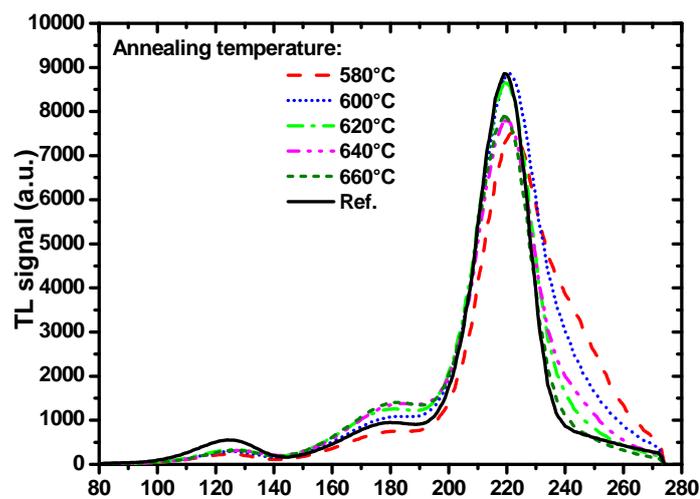

Fig. 7. TL glow-curves typical for all groups of detectors and durations of HTGA annealing in the range of temperature 580-660°C.

Analysis of the area under the main dosimetric peak (i.e. without the high temperature 'tail') indicates the total recovery of sensitivity after annealing at temperatures 600°C and 620°C, both for 20 and 30 minutes. However, taking into account the contribution of the high temperature 'tail' of the main peak, the best results of its reduction were obtained for the highest temperature used (i.e. 660°C) for annealing times of at least 20 minutes (10 minutes apparently was not long enough to result in full recovery of the shape of the glow-curve). Figures 8 and 9 show the results for 20 minutes of HTGA annealing time for all annealing temperatures applied (580-660°C).

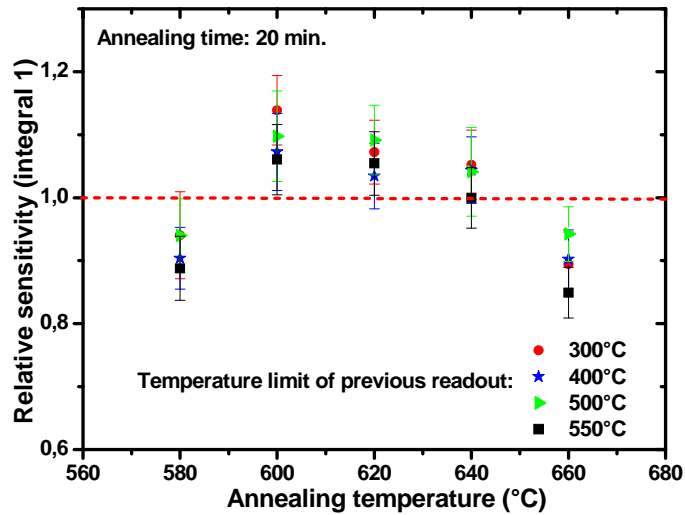

Fig. 8. The relative sensitivity of the main peak area for 20 minutes of HTGA annealing in the specific temperature from the range 580°C-660°C.

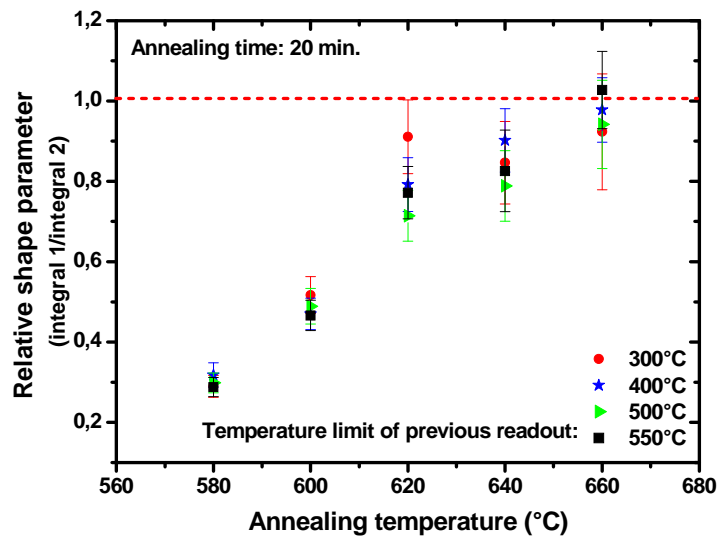

Fig. 9. The TL relative 'shape' parameter (integral 1/integral 2) after 20 minutes of HTGA annealing in the specific temperature from the range 580°C-660°C.

In addition, the results for groups of detectors readout previously up to a different temperature limit depend differently on the annealing time; see Fig. 10. Results confirm that when it comes to reduction of the high temperature 'tail' of the main peak, the temperatures 580°C and 600°C give the worst results. For these temperatures the dependence on duration of annealing is different than for 640°C and 660°C: in the former case prolongation of annealing time is unfavourable as the high temperature 'tail' of the main peak increases, whereas for higher annealing temperatures the annealing gives better effects when it is longer. This trend is indicated with arrows in Fig. 10. At the temperature of 620°C the recovery of sensitivity is at the similar level for all annealing duration.

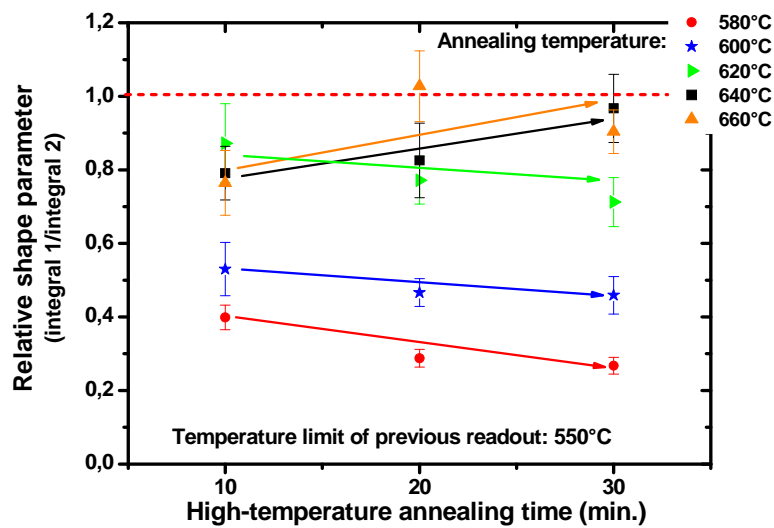

Fig. 10. The relative 'shape' parameter (integral 1/integral 2) for group of detectors readout previously up to 550°C (arrows show general trends).

Results obtained by deconvolution of the glow-curve into five individual peaks (an example of a deconvoluted glow-curve is presented in Figure 11) are similar to those obtained by integrating the glow-curves in the respective temperature ranges. Namely, the peak 4 area is the closest to the reference group after HTGA annealing at 620°C and 640°C, whereas the area under the peak 5 is responsible for the occurrence of the characteristic 'tail' at the high temperature and is effectively eliminated by the highest temperature of annealing. The examples of results are presented in Figures 12 and 13.

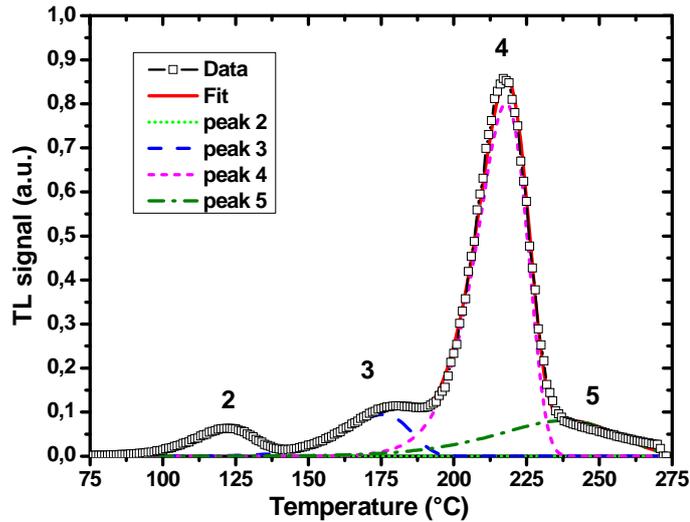

Fig. 11. An example of the glow-curve resulted from the HTGA annealing at 600°C for 10 min. followed by 240°C for 10 min. deconvoluted into five peaks.

Previous analysis leads to the conclusion that the temperature around 600°C and 620°C effectively regenerates the area of the main peak, i.e. the intensity of the dosimetric peak, but does not reduce the high temperature 'tail' fully, while higher temperature of annealing better reduce the growth of high-temperature 'tail', but reduces also the intensity of the main dosimetric peak.

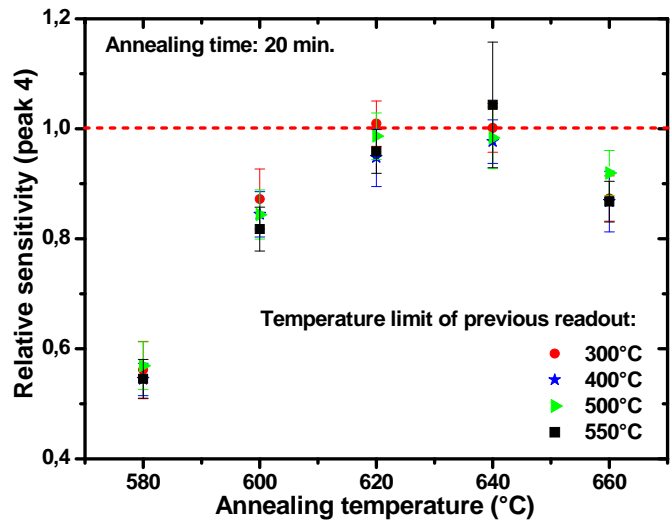

Fig. 12. The relative sensitivity of the area under the deconvoluted peak 4 for 20 minute HTGA annealing as a function of HTGA annealing temperature with a distinction between groups of detectors readout previously up to different temperature limits.

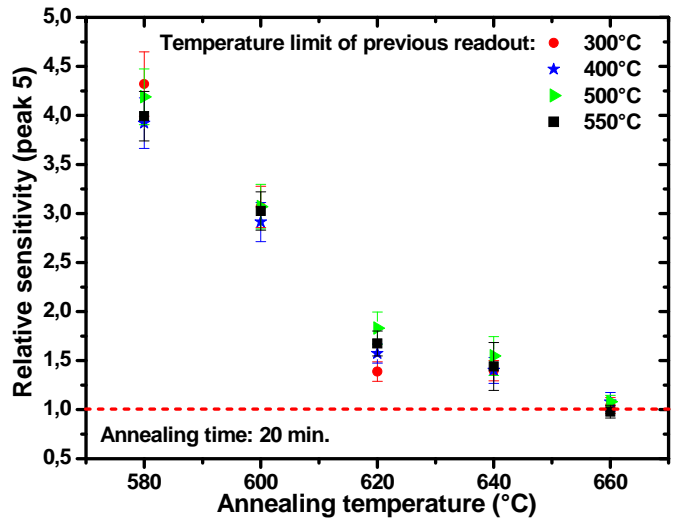

Fig. 13. The relative sensitivity of the area under the deconvoluted peak 5 for 20 minute HTGA annealing as a function of HTGA annealing temperature with a distinction between groups of detectors readout previously up to different temperature limits.

Thus, in the last step of searching for the best annealing procedure to be able to recover thermally-induced sensitivity loss, the impact on the high-temperature 'tail' of the glow-curve was examined by adding 270°C/10 min. annealing to the annealing procedure (stage (7)). A similar technique was applied by Tang et al. (2000). Improving the shape of the glow-curve would give the possibility of annealing at lower temperatures (e.g. 600°C, 620°C) with higher sensitivity in the main peak preserved.

Additional annealing (270°C/10min.) proved to be an effective tool for reducing the high temperature 'tail' of the glow-curve. The intensity of the main peak is the largest after annealing in 600°C; however, it appears that annealing at 620°C restored the glow-curve shape fully (Fig. 14). In the case of lower temperatures (i.e. 600°C) a residual of the high temperature 'tail' is still visible, and for the highest annealing temperature (640°C) the area under the peak 5, responsible for the high temperature 'tail' is smaller (up to 40%) than in the reference group (Fig. 15). The highest dispersion of responses for each group of detectors and the uncertainty of the values was obtained for 10 minutes of high-temperature annealing.

Finally, taking into account both qualitative and quantitative analysis, which was carried out in two ways (TL glow-curve integration within certain ranges and deconvolution of the glow-curve), annealing conditions have been established (temperature and time) that effectively remove a decrease of the sensitivity of MCP detectors due to the high temperature readout. Due to the very high intensity of the main peak, the effective reduction of the high temperature 'tail' and a relatively small dispersion indicated between groups of detectors, the HTGA annealing at 620°C for 20 minutes has been selected, followed by 270°C/10 min. and 240°C/10 min., in order to restore the original shape of the glow-curve.

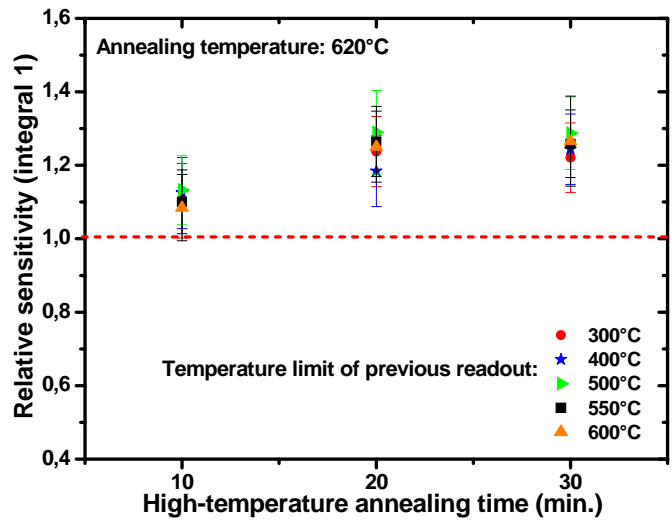

Fig. 14. The relative sensitivity of the main peak integral (integral 1) for annealing temperature 620°C as a function of annealing duration, distinguishing different groups of detectors.

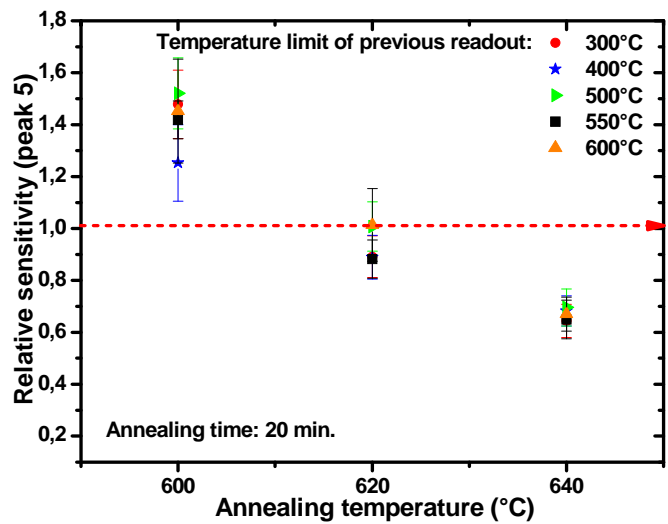

Fig. 15. The relative sensitivity of the area of the deconvoluted peak 5 of the glow-curve for 20 minutes annealing duration as a function of annealing temperature, distinguishing different groups of detectors.

### 4.3. Effect of high-temperature annealing on radiation & thermally-induced sensitivity recovery

After applying selected annealing procedure, the recovery of thermally-induced sensitivity loss of detectors used in high-dose gamma experiment (exposed with nine steps of doses up to 500 kGy) was expected. The loss in sensitivity was caused by high temperature limit of readout, the application of which was necessary to record the peak 'B'; however, high doses also damage the sensitivity of detectors. It was also decided to prolong HTGA annealing to 30 minutes, and then to 60 minutes in order to regain the rest of the sensitivity which was not recovered after the first 20 minute HTGA procedure (for detectors previously irradiated with 50 kGy gamma doses and higher). HTGA annealing was always followed by annealing at 270°C for 10 minutes and 240°C for 10 minutes. The resulted relative sensitivity recovery of MCP detectors after subsequent annealing conditions is shown in Fig. 16. Due to lack of not used detectors of this batch (produced in 2005) the result for the group exposed to lowest dose and treated routinely was adopted as the reference value.

The HTGA annealing procedure at 620°C extended to 60 minutes followed by annealing at 270°C for 10 minutes and at 240°C for 10 minutes was tested for the detectors from the 10 MeV electron beam exposures at doses of 5 kGy, 20 kGy and 100 kGy. Their sensitivity damage relative to the reference group (not used earlier detectors of the same production batch) tested before the HTGA annealing is shown in Fig. 17. Fig.17 also presents recovery of relative TL sensitivity of these detectors resulting from the HTGA 60 minutes procedure. Full recovery of the sensitivity of detectors was achieved.

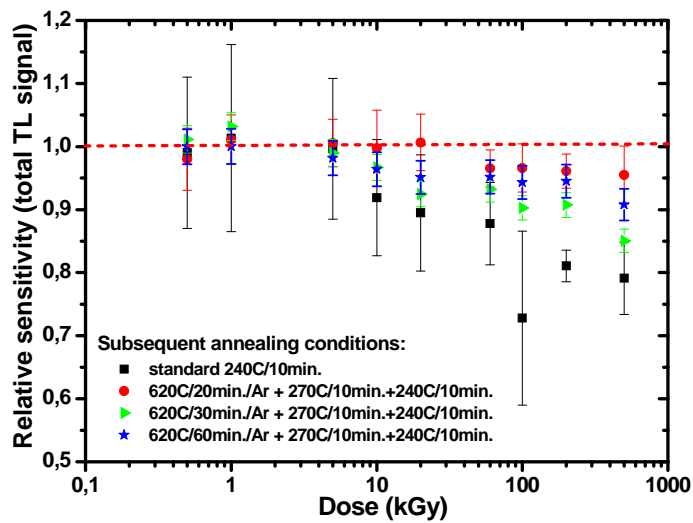

Fig. 16. The relative sensitivity of the detectors from high-dose gamma experiment after different subsequent annealing conditions.

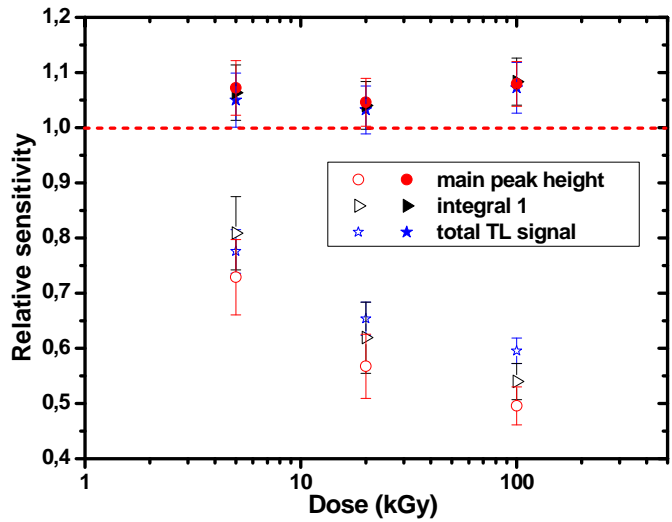

Fig. 17. The relative sensitivity of the detectors from high-dose electron experiment annealed: in standard conditions – empty symbols, with the special HTGA procedure
(620°C/20 min./Ar+270°C/10 min.+240°C/10 min.) – filled symbols.

**5. Summary and conclusions**

LiF:Mg,Cu,P can measure doses ranging from below 1 µGy to about 1 MGy (twelve orders of magnitude), also in mixed radiation fields with thermal and epithermal neutron presence; however, up to now it was not possible to reuse detectors after high-dose measurement due to their sensitivity loss.

To recover sensitivity of MCP detectors several annealing procedures were investigated, applying different temperatures (from 400°C up to 700°C) for different periods of time (10-60 minutes) in argon atmosphere. We have found that temperatures close to 620°C allow the restoration of TL sensitivity of MCPs, which have undergone earlier high temperature readout. This procedure has been applied to the MCP samples which have been earlier irradiated with high radiation doses and readout up to 600°C. It allowed regaining part of the sensitivity lost due to the non-standard readout process. The next step of our study was finding out a thermal treatment procedure which also allowed regaining the TL sensitivity damage caused by high radiation doses.

We have proved that the loss of MCP sensitivity, as a consequence of high-temperature readout can be fully recovered by the HTGA 620°C/20 min. annealing, followed by 270°C/10 min. and 240°C/10 min., while sensitivity loss caused by high-dose measurements (combined high-dose and high-temperature sensitivity loss) after gamma and electron high-dose irradiation can be fully restored by the HTGA 620°C/1h annealing, followed by 270°C/10 min. and 240°C/10 min. for the samples exposed earlier to doses up to $10^5$ Gy.

In this way we were able to recover MCP sensitivity fully, allowing for reuse of the samples after high-dose and high-temperature treatment. We have proven that the thermally and radiation-induced sensitivity loss of LiF:Mg,Cu,P is fully reversible. This indicates that high temperature and high doses of radiation are not causing irreversible changes in the material so whatever changes or damages are occurring in the defect structure (trapping, luminescent and competitive centres) of this phosphor, it can be restored by a certain sequence of annealing procedures. The work can therefore be of potential practical importance and has extended our phenomenological understanding of the behaviour of this material following high-dose and high-temperature treatment.

**Acknowledgments**

Work performed within the strategic research project 'Technologies supporting the development of safe nuclear power' financed by the National Centre for Research and Development (NCBiR). Research Task 'Research and development of techniques for the controlled thermonuclear fusion', Contract No. SP/J/2/143234/11.